%% file: 14665.tex
\documentclass[longauth,traditabstract,letter]{aa}
\usepackage{graphicx}
\usepackage{txfonts}
\usepackage{natbib}
\bibpunct{(}{)}{;}{a}{}{,}
\usepackage{hyperref}
\usepackage{ucs}
\usepackage[utf8x]{inputenc}
\long\def\beginpgfgraphicnamed#1#2\endpgfgraphicnamed{\includegraphics{#1}}
\newcommand{\garradd}{C/2008 Q3 (Garradd)}
\newcommand{\xne}{x_{n_\mathrm{e}}}
\newcommand{\rh}{r_\mathrm{h}}
\newcommand{\range}{1.7\textrm{--}2.8}

\newcommand{\transi}{$1_{10}$--$1_{01}$}
\newcommand{\transii}{$2_{12}$--$1_{01}$}
\newcommand{\transiii}{$1_{11}$--$0_{00}$}
\newcommand\water{\ifmmode{\rm H_2O}\else H$_2$O\fi}
\newcommand\kms{\ifmmode{\rm km\thinspace s^{-1}}\else km\thinspace s$^{-1}$\fi}
\newcommand\ms{\ifmmode{\rm m\thinspace s^{-1}}\else m\thinspace s$^{-1}$\fi}
\newenvironment{DIFnomarkup}{}{}
\graphicspath{{./figures/}}

\begin{document}

\title{HIFI observations of water in the atmosphere of comet C/2008 Q3
(Garradd)\thanks{{\it Herschel} is an ESA space observatory with science
instruments provided by European-led Principal Investigator consortia
and with important participation from NASA.}}

\author{P.~Hartogh\inst{\ref{inst1}}
  \and J.~Crovisier\inst{\ref{inst2}}
  \and M.~de~Val-Borro\inst{\ref{inst1}}
  \and D.~Bockel\'ee-Morvan\inst{\ref{inst2}}
  \and N.~Biver\inst{\ref{inst2}}
  \and D.~C.~Lis\inst{\ref{inst3}}
  \and R.~Moreno\inst{\ref{inst2}}
  \and C.~Jarchow\inst{\ref{inst1}}
  \and M.~Rengel\inst{\ref{inst1}}
  \and M.~Emprechtinger\inst{\ref{inst3}}
  \and S.~Szutowicz\inst{\ref{inst4}}
  \and M.~Banaszkiewicz\inst{\ref{inst4}}
  \and F.~Bensch\inst{\ref{inst5}}
  \and M.~I.~Blecka\inst{\ref{inst4}}
  \and T.~Cavali\'e\inst{\ref{inst1}}
  \and T.~Encrenaz\inst{\ref{inst2}}
  \and E.~Jehin\inst{\ref{inst6}}
  \and M.~K\"uppers\inst{\ref{inst7}}
  \and L.-M.~Lara\inst{\ref{inst8}}
  \and E.~Lellouch\inst{\ref{inst2}}
  \and B.~M.~Swinyard\inst{\ref{inst9}}
  \and B.~Vandenbussche\inst{\ref{inst10}}
  \and E.~A.~Bergin\inst{\ref{inst11}}
  \and G.~A.~Blake\inst{\ref{inst3}}
  \and J.~A.~D.~L.~Blommaert\inst{\ref{inst10}}
  \and J.~Cernicharo\inst{\ref{inst12}}
  \and L.~Decin\inst{\ref{inst10},\ref{inst23}}
  \and P.~Encrenaz\inst{\ref{inst13}}
  \and T.~de~Graauw\inst{\ref{inst18},\ref{inst24},\ref{inst14}}
  \and D.~Hutsemekers\inst{\ref{inst6}}
  \and M.~Kidger\inst{\ref{inst15}}
  \and J.~Manfroid\inst{\ref{inst6}}
  \and A.~S.~Medvedev\inst{\ref{inst1}}
  \and D.~A.~Naylor\inst{\ref{inst16}}
  \and R.~Schieder\inst{\ref{inst17}}
  \and N.~Thomas\inst{\ref{inst19}}
  \and C.~Waelkens\inst{\ref{inst10}}
  \and P.~R.~Roelfsema\inst{\ref{inst18}}
  \and P.~Dieleman\inst{\ref{inst18}}
  \and R.~G\"usten\inst{\ref{inst20}}
  \and T.~Klein\inst{\ref{inst20}}
  \and C.~Kasemann\inst{\ref{inst20}}
  \and M.~Caris\inst{\ref{inst20}}
  \and M.~Olberg\inst{\ref{inst21},\ref{inst18}}
  \and A.~O.~Benz \inst{\ref{inst22}}
  }

\institute{
  Max-Planck-Institut f\"ur Sonnensystemforschung, 37191
    Katlenburg-Lindau, Germany\label{inst1}
  \and LESIA, Observatoire de Paris, 5 place Jules Janssen, 92195
  Meudon, France\label{inst2}
  \and California Institute of Technology, Pasadena, CA 91125, USA\label{inst3}
  \and Space Research Centre, Polish Academy of Sciences, Warsaw,
    Poland\label{inst4}
  \and DLR, German Aerospace Centre, Bonn-Oberkassel, Germany\label{inst5}
  \and Institute d'Astrophysique et de Geophysique, Université de Liège,
    Belgium\label{inst6}
  \and Rosetta Science Operations Centre, European Space Astronomy
    Centre, European Space Agency, Spain\label{inst7}
  \and Instituto de Astrof\'isica de Andaluc\'ia (CSIC), Spain\label{inst8}
  \and Space Science \& Technology Department, Rutherford Appleton
    Laboratory, UK\label{inst9}
  \and Instituut voor Sterrenkunde, Katholieke Universiteit Leuven,
  Belgium\label{inst10}
  \and Astronomy Department, University of Michigan, USA\label{inst11}
  \and Laboratory of Molecular Astrophysics, CAB-CSIC, INTA, Spain\label{inst12}
  \and Sterrenkundig Instituut Anton Pannekoek, University of Amsterdam,
  Science Park 904, NL-1098 Amsterdam, The Netherlands\label{inst23}
  \and LERMA, Observatoire de Paris, France\label{inst13}
  \and SRON Netherlands Institute for Space Research, Landleven 12,
  9747 AD, Groningen, The Netherlands\label{inst18}
  \and Leiden Observatory, University of Leiden, the Netherlands\label{inst24}
  \and Joint ALMA Observatory, Chile\label{inst14}
  \and {\it Herschel} Science Centre, European Space Astronomy
  Centre, European Space Agency, Spain\label{inst15}
  \and Department of Physics and Astronomy, University of Lethbridge,
  Canada\label{inst16}
  \and 1st Physics Institute, University of Cologne, Germany\label{inst17}
  \and Physikalisches Institut, University of Bern, Switzerland\label{inst19}
  \and Max-Planck-Institut f\"ur Radioastronomie, Auf dem H\"ugel 69, 53121
  Bonn, Germany\label{inst20}
  \and Chalmers University of Technology, SE-412 96 Göteborg, Sweden\label{inst21}
  \and Institute of Astronomy, ETH Z\"urich, 8093 Z\"urich, Switzerland\label{inst22}
  }

\date{Received March 31, 2010; accepted }

\abstract{ High-resolution far-infrared and sub-millimetre spectroscopy
of water lines is an important tool to understand the physical and
chemical properties of cometary atmospheres.
We present  observations of several rotational ortho- and
para-water transitions in comet \garradd{} performed with HIFI on
{\it Herschel}. These observations have provided the first detection of
the $2_{12}$--$1_{01}$ (1669 GHz) ortho and $1_{11}$--$0_{00}$
(1113 GHz) para transitions of water in a cometary spectrum. In
addition, the ground-state transition $1_{10}$--$1_{01}$ at
557 GHz is detected and mapped.  By detecting several water lines
quasi-simultaneously and mapping their emission we can constrain
the excitation parameters in the coma. Synthetic line profiles are
computed using excitation models which include excitation by
collisions, solar infrared radiation, and radiation trapping. We obtain the
gas kinetic temperature, constrain the electron density profile, and
estimate the coma expansion velocity by analyzing the map and line
shapes. We derive water production rates of $\range \times10^{28}\
\mathrm{s}^{-1}$ over the range $r_\mathrm{h}=1.83$--$1.85$ AU.}


\maketitle

\section{Introduction}

Comets spend most of their lifetime in the outer Solar System and
therefore have not undergone considerable thermal processing.  Line
emission is useful to study the physical and chemical conditions of
cometary atmospheres, and their relation to other bodies in the Solar
System \citep{2002EM&P...90..323B,2004come.book..391B}.  

Water molecules in cometary atmospheres are excited due to collisions
with other molecules and radiative pumping of the fundamental
vibrational levels by the solar infrared flux. The \transi{} ortho-water
transition at 557 GHz is one of the strongest lines in cometary comae,
but it cannot be detected directly from the ground due to absorption in
the Earth's atmosphere \citep{1987A&A...181..169B}. Water vapour
production has been estimated previously from the ground through
measurements of its photodissociation product, the OH radical \citep[see
e.g.][]{1982come.coll..433A}, and water high vibrational bands
\citep{2004come.book..391B}.
The \transi{} rotational transition of ortho-water at 557 GHz has been
observed using heterodyne techniques by the Submillimeter Wave
Astronomical Satellite (SWAS)
\citep{2000ApJ...539L.151N,2001Icar..154..345C}, and later with Odin
\citep{2003A&A...402L..55L,2007P&SS...55.1058B,2009A&A...501..359B}.

ESA's {\it Herschel} Space Observatory was successfully launched on May 14,
2009 and entered a Lissajous orbit around the $L_2$ Lagrangian point
\citep{2010Herschel}. The Heterodyne Instrument for the Far-Infrared
(HIFI) onboard {\it Herschel} has continuous coverage in five frequency bands
in the 480--1150 GHz range, and an additional dual frequency band
covering the 1410--1910 GHz range that are not observable from the
ground \citep{2010HIFI}.  HIFI's submillimetre frequency coverage is of
great importance to observe water vapour in Solar System objects such as
cometary comae with unique sensitivity and the required high spectral
resolution to resolve the emission lines.

Comet \garradd{} is a long-period comet ($P = 1.9\times10^5$)
with a highly eccentric orbit ($e=0.99969$). It passed perihelion on
June 23, 2009 at a distance of 1.7982 AU from the Sun and was observed
with HIFI in July 2009 as part of the {\it Herschel} guaranteed time key
project ``Water and related chemistry in the Solar System''
\citep{2009P&SS...57.1596H}.  In this letter, we describe the
observations of water and the analysis of part of the data set. Water
production rates derived from our radiative transfer models are
presented.

\section{{\it Herschel} HIFI observations}
\label{sec:observations}

Comet \garradd{} was observed with HIFI about one month post-perihelion
in the period July 20--27 2009, when the comet was at $\rh \simeq 1.8$
AU from the Sun and $\Delta \simeq 1.9$ AU from {\it Herschel}
(Table~\ref{table:1}). HIFI's submillimetre high-resolution heterodyne
spectrometer is designed to have noise levels close to the quantum limit
\citep{2010HIFI}. The spectral resolution is 1.1 MHz and 140 KHz
provided by the Wideband Spectrometer (WBS) and the High Resolution
Spectrometer (HRS), enabling HIFI to resolve spectrally cometary line
shapes.

\begin{table*}
  \caption{Summary of HIFI observations of comet \garradd{} and retrieved water
  production rates. The observations were conducted using standard frequency
  switching (FSw), position switching (PSw) and on-the-fly spectral maps (OTF)
  observing modes.  The error bars in line intensity, velocity shift and
  production rate are the statistical errors.
}
  \begin{DIFnomarkup} 
  \label{table:1} \centering
  \begin{tabular}{c c c c c c c c c r@{}l c}
    \hline\hline Obs.\ ID & UT start date & $\Delta$ & $r_\mathrm{h}$ & Band &
    Obs.\ mode & $\nu_{ij}$ & Integration & Intensity & 
    \multicolumn{2}{c}{Velocity shift} & $Q_\water{}$ \\
    & [mm/dd.ddd] & [AU] & [AU] & & & [GHz] & [s] & [K \kms] & 
    \multicolumn{2}{c}{[\ms]} & [$10^{28} \mathrm{s}^{-1}$] \\
    \hline
    \input{obs_14665.txt}
    \hline
  \end{tabular}
  \end{DIFnomarkup} 
\end{table*}

We present a summary of the HIFI observations in Table~\ref{table:1}.
These HIFI observations constitute the first detection of the
$2_{12}$--$1_{01}$ transition of ortho-water at 1669.9 GHz, and the
$1_{11}$--$0_{00}$ transition of para-water at 1113.3 GHz in cometary
coma. In addition, the ground-state transition $1_{10}$--$1_{01}$ at 557
GHz is detected. Multi-line observations can provide more reliable water
production rates using molecular excitation codes.

The data were acquired using the standard frequency switching and
position switching observing modes. In the latter mode, a reference cold
sky position separated from the comet by $0.5\degr$ was observed for the
same amount of time and subtracted from the on-source observations.
Three additional on-the-fly (OTF) maps were obtained at 557 GHz with the
standard observational modes (the OTF mode was not released at that
time).  The {\it Herschel} telescope has a diameter of 3.5 m, with a
corresponding HIFI Half Power Beam Width (HPBW) of $12.7\arcsec$,
$19.2\arcsec$, and $38.1\arcsec$ at 1669, 1113, and 557 GHz,
respectively.  This represents projected distances of $1.7$--$5.1
\times10^{4}$ km at the comet. The JPL's HORIZONS
system\footnote{\url{http://ssd.jpl.nasa.gov/?horizons}} is used to
calculate the ephemerides and the relative motion of the comet with
respect to the satellite.

The data were reduced to a level 2 product using the standard {\it Herschel}
Interactive Processing Environment (HIPE) routines for HIFI
\citep{2010HIFI}.  The frequency scale of the observed spectra was
corrected for the geocentric velocity of the comet and the spacecraft
orbital velocity.  We scaled the main beam brightness temperature using
the beam efficiency of the {\it Herschel} telescope, ranging between
0.65--0.75 at different frequencies.  Integrated line intensities and
velocity offsets in the comet rest frame are given in
Table~\ref{table:1}.

\section{Data analysis}
\label{sec:analysis}

\subsection{Water line emission}

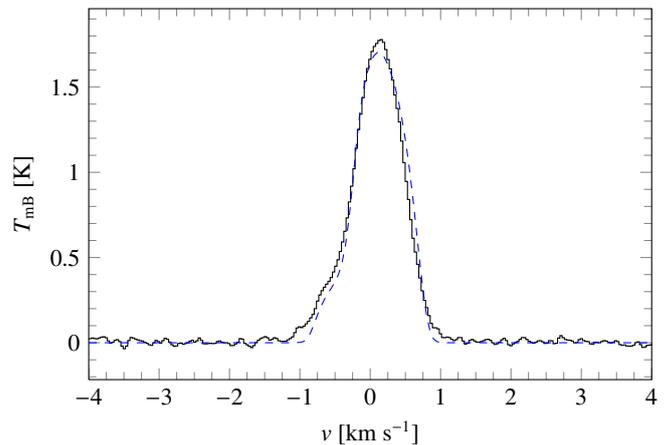
\begin{figure}
  \centering
  \begin{DIFnomarkup} 
  \beginpgfgraphicnamed{14665fg1}
    \begin{axis}[xlabel={$v$ [km s$^{-1}$]},
      ylabel={$T_\mathrm{mB}$ [K]}, xmin=-4, xmax=4, height=6.5cm, width=\hsize,
      minor x tick num=3, minor y tick num=4]
      \addplot[mark=none,const plot mark right]
      table[x index=0,y index=1] 
      {/home/miguel/HssO/Garradd/paper/fig1.txt};
      \addplot[mark=none,blue,dashed]
      table[x index=0,y index=1] 
      {/home/miguel/HssO/Garradd/paper/fig1-model.txt};
    \end{axis}
  \endpgfgraphicnamed
  \end{DIFnomarkup} 
  \caption{Spectrum of the $1_{10}$--$1_{01}$ ortho-water line at 556.936 GHz in
  comet \garradd{} obtained by the HRS on July 20.94 UT in FSw mode. The
  velocity scale is given with respect to the comet rest frame.  The dashed
  line shows a synthetic profile computed with the Monte Carlo model for
  isotropic outgassing at $v_\mathrm{exp}$ = 0.55 \kms, $T = 15$ $\mathrm{K}$,
  and $\xne{} = 0.2$.}
  \label{fig:557}
\end{figure}

Figure~\ref{fig:557} shows the spectrum of the \transi{} ortho-water line,
corresponding to a long integration in frequency switching mode with a
throw of 18 MHz. Here the signal-to-noise ratio is very high. This line
is optically thick and slightly asymmetric due to self-absorption
effects in the coma. From the width of the non-self-absorbed redshifted
side of the line we obtain an estimate of 0.55 \kms\ for the coma
expansion velocity.  Values in the range $v_\mathrm{exp} = 0.5-0.8\
\kms$ are typical for relatively weak comets with total water production
rate $Q_\water{} < 10^{29}\ \mathrm{s}^{-1}$ at 
$r_\mathrm{h} > 1.2\ \mathrm{AU}$.

\begin{figure}
  \centering
  \begin{DIFnomarkup} 
  \beginpgfgraphicnamed{14665fg2}
    \begin{axis}[xlabel={$v$ [km s$^{-1}$]},
      ylabel={$T_\mathrm{mB}$ [K]}, xmin=-4, xmax=4, height=6.5cm, width=\hsize,
      minor x tick num=3, minor y tick num=4]
      \addplot[mark=none,const plot mark right]
      table[x index=0,y index=1] 
      {/home/miguel/HssO/Garradd/paper/fig2.txt};
    \end{axis}
  \endpgfgraphicnamed
  \end{DIFnomarkup}
  \caption{Average of the HRS frequency-switched observations of the
  $1_{11}$--$0_{00}$ para-water line at 1113.343 GHz in comet
  \garradd{} obtained on July 22.34 and 27.78 UT. 
  }
  \label{fig:1113}
\end{figure}
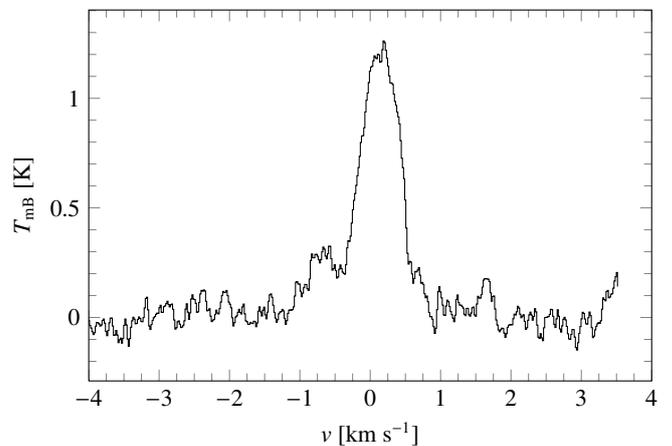

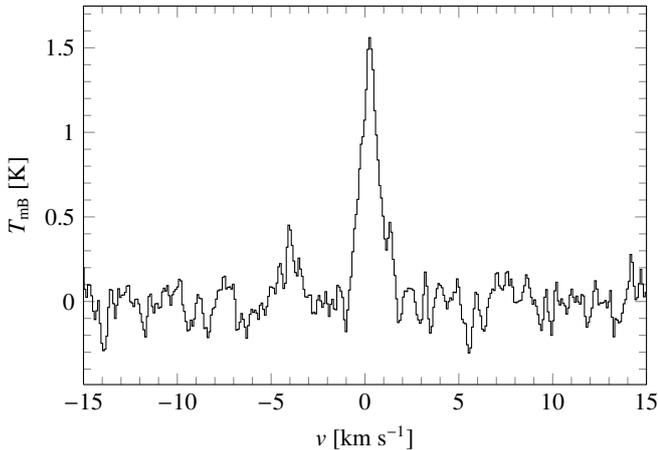
\begin{figure}[t]
  \centering
  \begin{DIFnomarkup} 
  \beginpgfgraphicnamed{14665fg3}
    \begin{axis}[xlabel={$v$ [km s$^{-1}$]},
      ylabel={$T_\mathrm{mB}$ [K]}, xmin=-15, xmax=15, height=6.6cm, width=\hsize,
      minor x tick num=4, minor y tick num=4]
      \addplot[mark=none,const plot mark right]
      table[x index=0,y index=1] 
      {/home/miguel/HssO/Garradd/paper/fig3.txt};
    \end{axis}
  \endpgfgraphicnamed
  \end{DIFnomarkup} 
  \caption{WBS frequency-switched observation of the $2_{12}$--$1_{01}$
   ortho-water line at 1669.905 GHz towards comet \garradd{} 
   obtained on July 27.73 UT.}
  \label{fig:1669}
\end{figure}

\begin{figure}
  \centering
  \resizebox{\hsize}{!}{
    \includegraphics[trim = 0 0 0 10mm, clip]{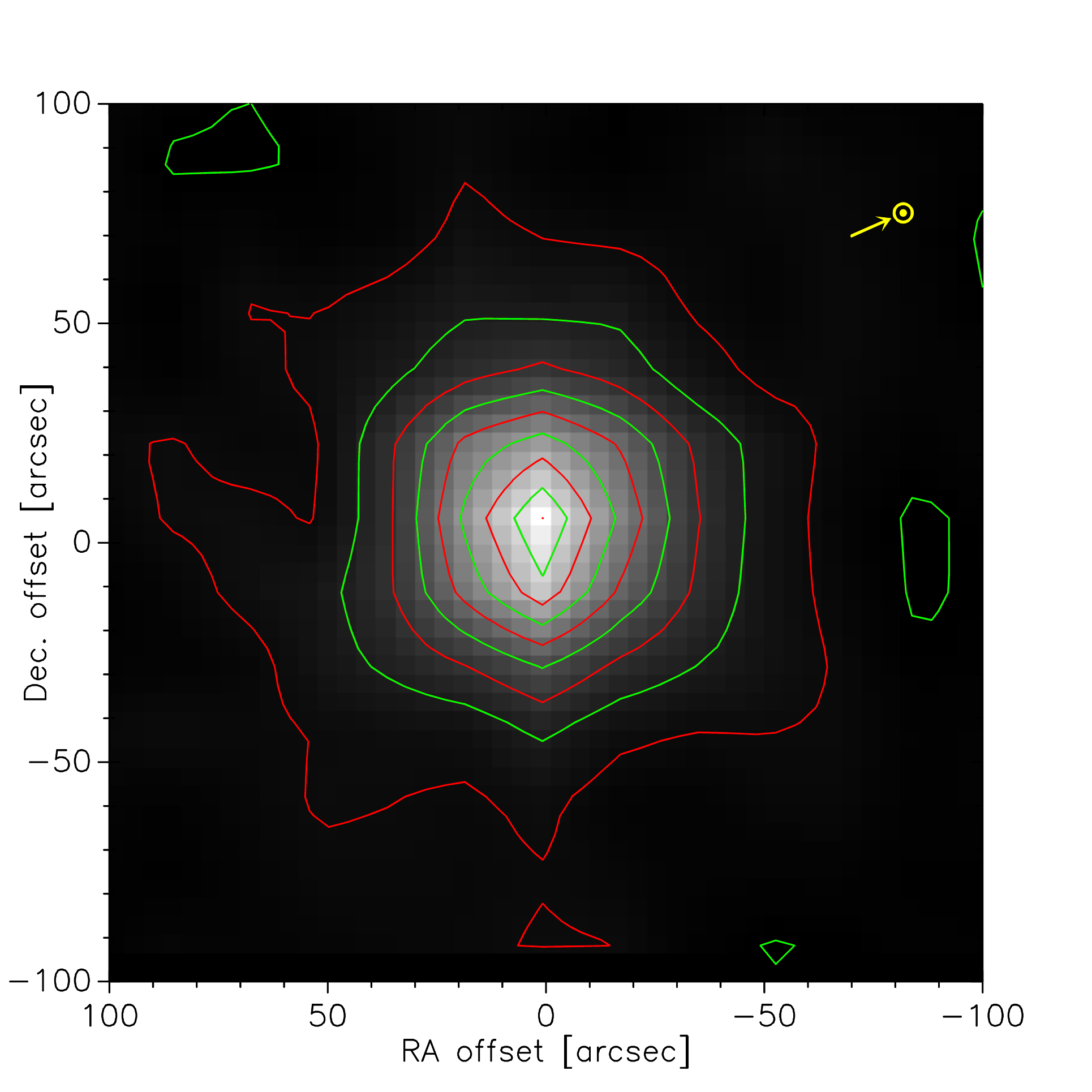}}
  \caption{On-the-fly map of the 557 GHz ortho line obtained by HRS on
  July 20.95 UT.  The contour step is 0.2 K \kms\ in brightness
  temperature, from 0 to 1.8 K \kms.  The map width is 3$\times$10$^{5}$
  km projected at the comet.  The direction towards the Sun is indicated
  by the arrow in the upper right corner.}
  \label{fig:OTM}
\end{figure}

Figure~\ref{fig:1113} shows the water spectrum of the \transiii{}
para-water line at 1113.3 GHz observed in frequency switching mode.
We averaged over two dates to increase the signal-to-noise ratio.
The \transii{} ortho-water line at 1669.9 GHz is detected in the HRS
and WBS data (Fig.~\ref{fig:1669}).
There are standing waves in the spectrum due to the frequency switching
observing mode.  The line intensity is strongly baseline-dependent.
Figure~\ref{fig:OTM} shows the on-the-fly integrated intensity map of
water emission at 557 GHz.
The outer regions beyond $\sim10\arcsec$ (13\,000 km)
from the nucleus are dominated by infrared fluorescence.

\subsection{Modelling}

We adopted the publicly available Accelerated Monte Carlo
radiative transfer code {\it ratran} \citep{2000A&A...362..697H}
to calculate water line emission in cometary coma. We used the
one-dimensional version of the code following the work of
\citet{2004ApJ...615..531B}. A water excitation model based on the
Sobolev escape probability method was also considered
\citep{1987A&A...181..169B,1997PhDT........51B}. These two approaches
provide very similar results \citep{2007A&A...473..303Z}.

The radial gas density profile for water was obtained using the standard
spherically symmetric Haser distribution \citep{1957BSRSL..43..740H}.
The expansion velocity is assumed to be constant in the coma. We use a constant
neutral gas kinetic temperature throughout the coma and the electron
temperature profile given by \citet{1997PhDT........51B} \citep[see
also][]{2004ApJ...615..531B}. We assume an ortho-to-para water abundance ratio
of 3 \citep[see e.g.,][]{1997Sci...275.1904C}.  
The few number of observed water lines, their non-simultaneity, and the
different beam sizes for each line prevented us to derive useful
constraints on the ortho-to-para ratio.
Molecular data for ortho- and para-water have been obtained from the
current version of the LAMDA database
\footnote{\url{http://www.strw.leidenuniv.nl/~moldata/}}
\citep{2005A&A...432..369S}.

The electron density profile is derived from the 1P/Halley in situ measurements
scaled to the water production rate and heliocentric distance of comet C/2008
Q3 (Garradd) \citep{1997PhDT........51B,2004ApJ...615..531B}. A scaling factor
to this electron density profile, $x_{n_\mathrm{e}}$, is introduced as a free
parameter to the model \citep{1997PhDT........51B}. Observations of the 557 GHz
line in comets with Odin have shown that the radial profiles of the line
brightness are best explained with $\xne{}=0.2$ \citep{2007P&SS...55.1058B}.
For the Monte Carlo code, the water-electron collision rates from
\citet{2004MNRAS.347..323F} are used.

Infrared pumping of vibrational bands by solar radiation contributes to the
excitation in the outer coma where the gas and electron densities are low
\citep{1987A&A...181..169B}. We use the effective pumping rates for the lowest
rotational levels of the ground vibrational state of ortho-water from
\citet{2007A&A...473..303Z}. Effective pumping rates for para-water were
computed as described in \citet{2007A&A...473..303Z}.

The relative level population of the lowest levels of para-water
is shown in {\it online} Fig.~\ref{fig:lp}. In
the inner coma, frequent collisions between water molecules lead to a
thermal equilibrium distribution, and electrons are cooled down close to
the neutral gas temperature \citep[e.g][]{2004ApJ...615..531B}. Given
the beam sizes (between 17\,000 and 51\,000 km projected on the comet),
molecules from the outer coma contribute to the detected emission. The
line excitation in this region is dominated by water-electron collisions
and infrared fluorescence.

Once the level populations are calculated, we solve the radiative
transfer equation along different lines of sight through the comet
atmosphere covering $2.5\times\mathrm{HPBW}$, and compute the beam
averaged emission at the distance of the comet. By comparing to the
observed line emission, we deduce the water production rate.

\onlfig{5}{
\begin{figure}
  \centering
   \resizebox{\hsize}{!}{
   \includegraphics{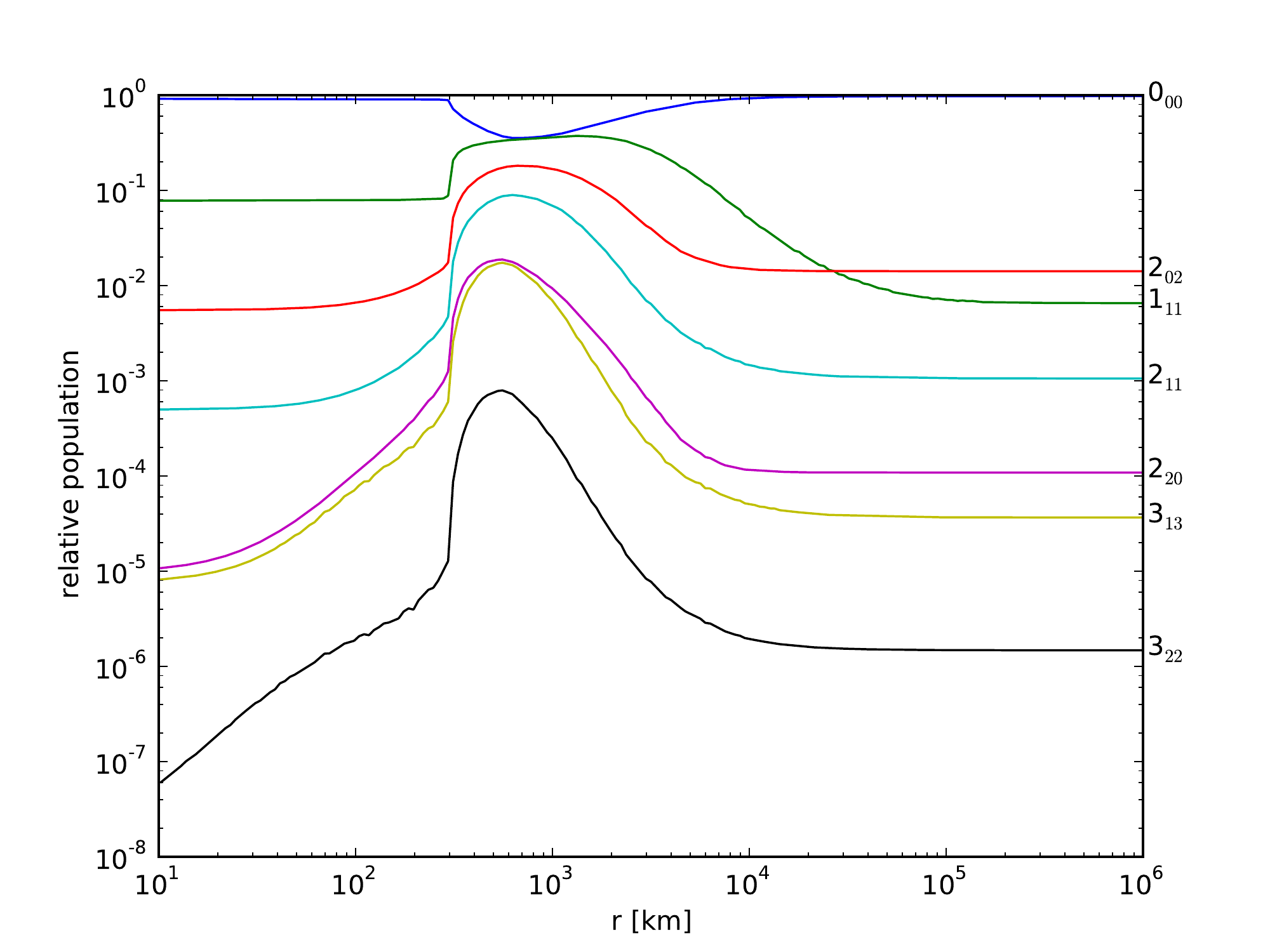}}
  \caption{
  Level population of para-water as a function of distance to
  the nucleus for $Q_\water = 1.7\times10^{28}\ \mathrm{s}^{-1}$, $\rh = 1.83$
  AU, $T = 15\ \mathrm{K}$, and $\xne{} = 0.2$.
  }
  \label{fig:lp}
\end{figure}
}

\subsection{Water production rates}

\begin{figure}
  \centering
  \resizebox{\hsize}{!}{
    \includegraphics[angle=270]{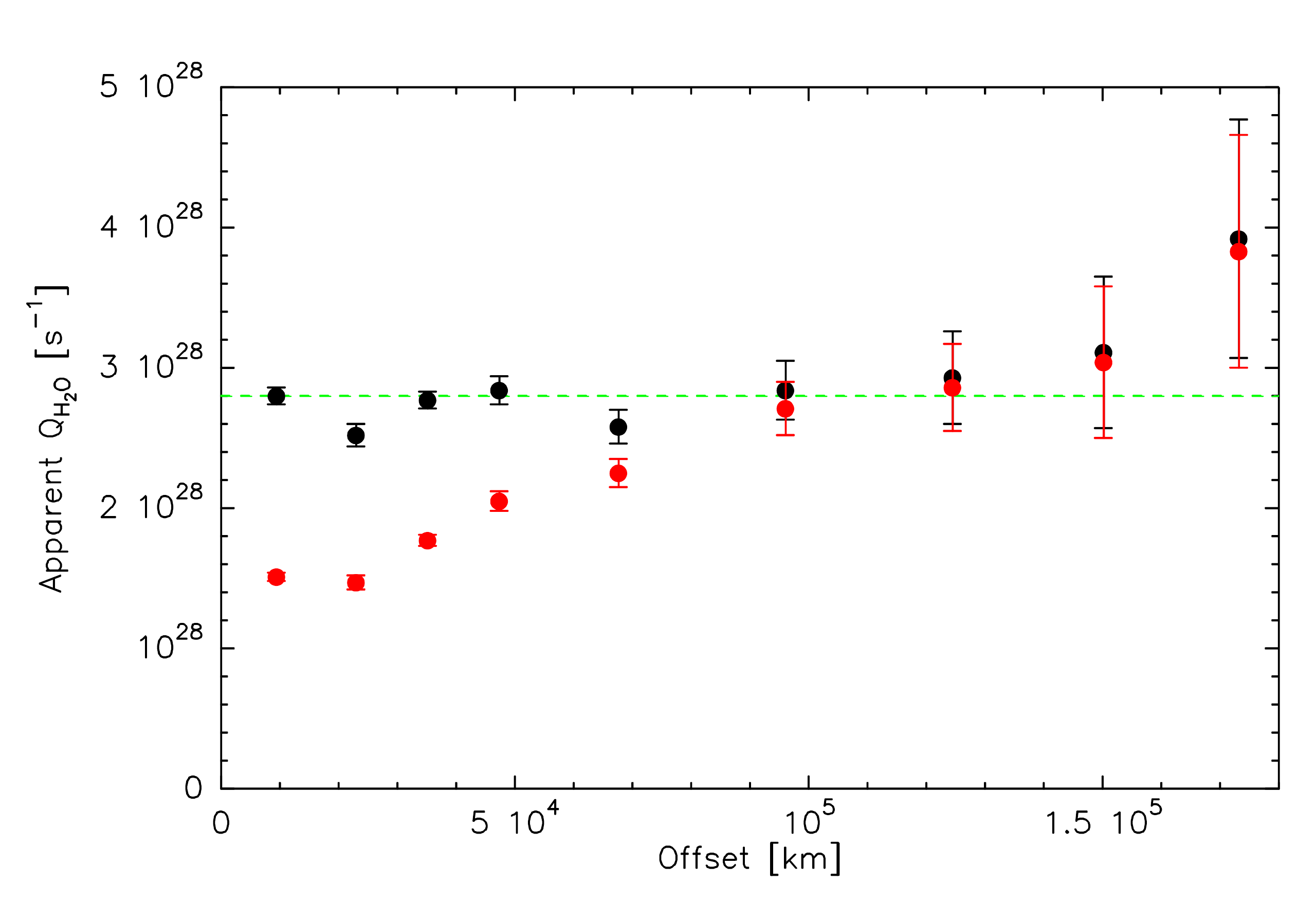}}
  \caption{Apparent water production rates as a function of offset deduced from
  the 557 GHz map shown in Fig.~\ref{fig:OTM}. Isotropic outgassing at
  $v_\mathrm{exp}$ = 0.55 \kms is assumed.  Results for $x_{n_\mathrm{e}}$ = 0.2
  and 1 are shown with black and red symbols, respectively.
  The dashed horizontal line shows the mean water production rate.
  }
  \label{fig:qh2o}
\end{figure}

Mapping observations shown in Fig.~\ref{fig:OTM} can be used to constrain the
neutral gas kinetic temperature and the electron density scaling factor
$x_{n_\mathrm{e}}$.  We constrained these model parameters by minimizing the
radial variation of the water production rate deduced from the intensity at 
different offset positions. Hence, deviations from the Haser law due, e.g., to
sublimating grains or structures induced by the rotation of the nucleus are
not considered. We used the HRS data with the two orthogonal
polarisations. We averaged radially the line intensity using bins of
$10\arcsec$ to 20\arcsec. In Fig.~\ref{fig:qh2o}, we show the retrieved
water production rate as a function of offset for an isotropic model
with an expansion velocity $v_\mathrm{exp}$ = 0.55 \kms, $T = 15$
$\mathrm{K}$, and $\xne{} = 0.2$ and 1.  We found that a constant
production rate is obtained for excitation parameters in the range $T =
15$--$25\ \mathrm{K}$ and $\xne{} =0.1$--$0.2$. The best-fit synthetic
line profile computed using the Monte Carlo code with $v_\mathrm{exp}
= 0.55\ \kms$, $T = 15\ \mathrm{K}$, and $\xne{} = 0.2$ is shown in
Fig.~\ref{fig:557}.  This model explains satisfactorily the observed
line shape.  A similar spectrum is obtained with the escape probability
method.  Both models differ by less than 5\% in the calculated line
intensities.

\begin{figure}[t]
  \centering
  \resizebox{\hsize}{!}{
    \includegraphics[angle=270]{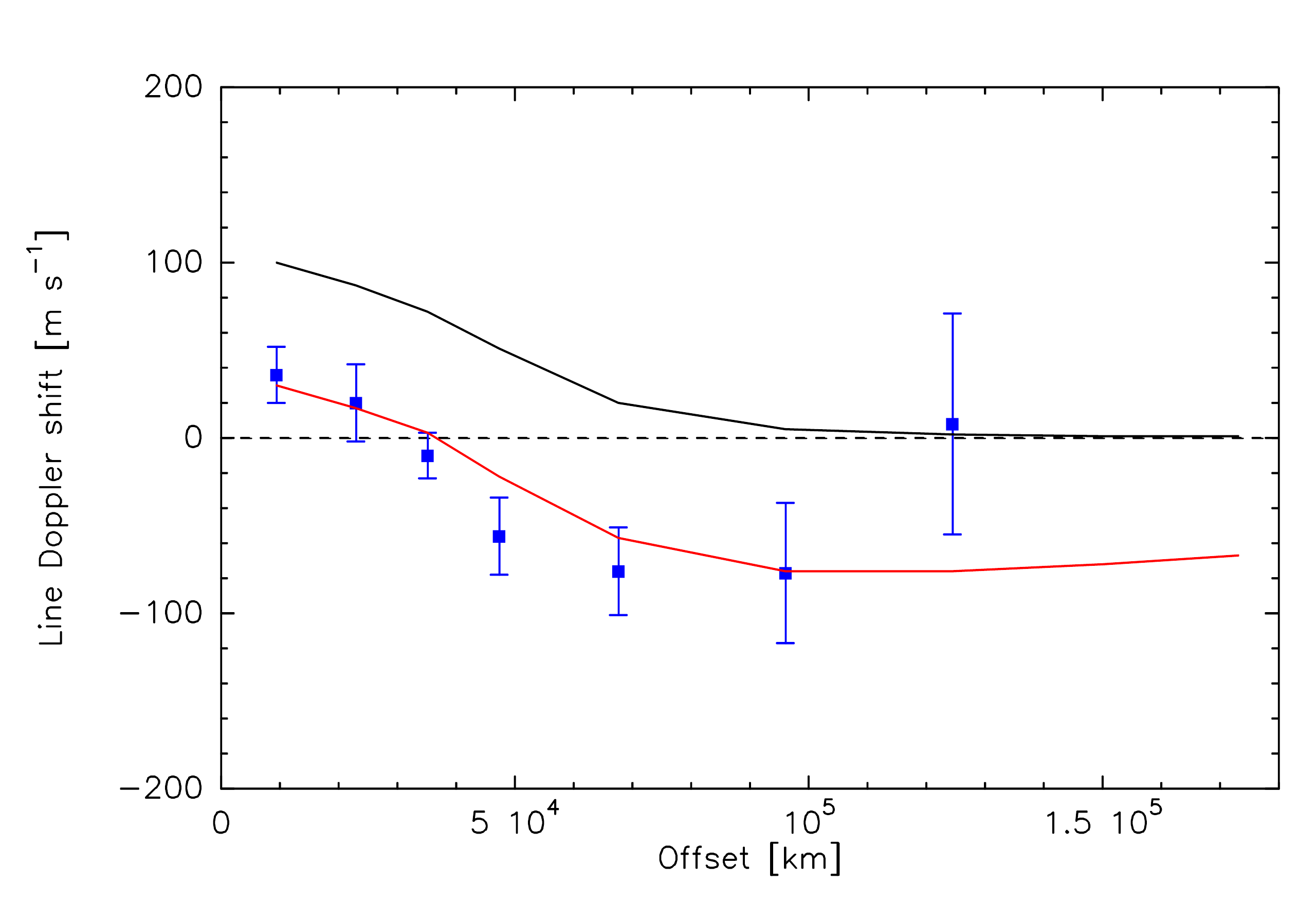}}
  \caption{557-GHz line Doppler shift as a function of position offset for
  models considering isotropic outgassing (black line) and anisotropic outgassing
  concentrated in a $\pm 130\degr$ cone (red line). The blue squares indicate the
  observed line Doppler shift.
  }
  \label{fig:vdop}
\end{figure}

We also investigated an anisotropic model with the outgassing restricted to a
$\pm 130\degr$ cone centered in the direction towards the Sun and
expansion velocity of 0.60 \kms. The anisotropic model provides a better
fit to the evolution of the line mean Doppler shift
(Fig.~\ref{fig:vdop}). This suggests a day/night asymmetry in the comet
outgassing which may affect production rate determinations.

The water production rates derived from different lines are in the range
$\range \times10^{28}\ \mathrm{s}^{-1}$. Those derived from the 1113 and 1669
GHz lines observed on the same day are consistent. A decrease of the comet
activity from July 20.9 to 27.8 UT is suggested. However, this needs to be
confirmed from a proper consideration of the outgassing asymmetry and possible
pointing offsets related to the comet ephemeris.

\section{Conclusions}
\label{sec:conclusions}

The {\it Herschel} Space Observatory provides unique new capabilities for the
detection of water in the Solar System.  HIFI's spectral range and high
resolution allowed for the direct detection of several water lines almost
simultaneously.  High spectral resolution is crucial to resolve the line shape
and asymmetries due to self-absorption.

On 20-27 July 2009, comet \garradd{} was observed with HIFI. The
high-resolution spectra of HIFI allows us to detect for the first time several
rotational water lines in cometary spectra.  A water production rate of $\sim
2\times10^{28}\ \mathrm{s}^{-1}$ was derived at heliocentric distance of 1.8 AU
using radiative transfer numerical codes which include collisional effects and
infrared fluorescence by solar radiation.

In future studies, HIFI will be able to detect water isotopes, and determine
the D/H ratio in active comets.

\begin{acknowledgements}
HIFI has been designed and built by a consortium of institutes and university
departments from across Europe, Canada and the United States under the
leadership of SRON Netherlands Institute for Space Research, Groningen, The
Netherlands and with major contributions from Germany, France and the US.
Consortium members are: Canada: CSA, U.Waterloo; France: CESR, LAB, LERMA,
IRAM; Germany: KOSMA, MPIfR, MPS; Ireland, NUI Maynooth; Italy: ASI, IFSI-INAF,
Osservatorio Astrofisico di Arcetri-INAF; Netherlands: SRON, TUD; Poland: CAMK,
CBK; Spain: Observatorio Astronómico Nacional (IGN), Centro de Astrobiología
(CSIC-INTA). Sweden: Chalmers University of Technology - MC2, RSS \& GARD;
Onsala Space Observatory; Swedish National Space Board, Stockholm University -
Stockholm Observatory; Switzerland: ETH Zurich, FHNW; USA: Caltech, JPL, NHSC.
HIPE is a joint development by the {\it Herschel} Science Ground Segment Consortium,
consisting of ESA, the NASA {\it Herschel} Science Center, and the HIFI, PACS and
SPIRE consortia.
This development has been supported by national funding agencies: CEA, CNES,
CNRS (France); ASI (Italy); DLR (Germany).
Additional funding support for some instrument activities has been provided by
ESA.
Support for this work was provided by NASA through an award issued by
JPL/Caltech.
LD  acknowledges financial support from the Fund for Scientific
Research - Flanders (FWO).
JB, LD, BV, CW acknowledge support from the Belgian Federal Science
Policy Office via the PRODEX Programme of ESA.
DCL is supported by the NSF, award AST-0540882 to the Caltech
Submillimeter Observatory.
SS and MB are supported by the Polish Ministry of Education and Science
(MNiSW).
LML thanks the Spanish MICIT through the project AyA 2009-08011-E.
\end{acknowledgements}

\bibliographystyle{aa}
\bibliography{references,ref,arxiv}

\Online

\end{document}

%% file: obs_14665.txt
1342180461 & 07/20.908 & 1.865 & 1.831 & 1a & FSw & 557 & 2181 & $1.594 \pm 0.009$ &             $+56\,\pm\,$ & 3 & $2.73 \pm 0.01$ \\
1342180462 & 07/20.933 & 1.865 & 1.831 & 1a & PSw & 557 & 381 & $1.717 \pm 0.023$ &             $+55\,\pm\,$ & 10 & $2.81 \pm 0.03$ \\
1342180463 & 07/20.936 & 1.865 & 1.831 & 1a & OTF & 557 & 2725 & & & & \\
1342180557 & 07/22.338 & 1.901 & 1.836 & 4b & FSw & 1113 & 2802 & $1.206 \pm 0.022$ &             $+24\,\pm\,$ & 13 & $1.82 \pm 0.03$ \\
1342180813 & 07/27.726 & 2.037 & 1.850 & 6b & FSw & 1669 & 3018 & $1.597 \pm 0.104$ &             $+283\,\pm\,$ & 28 & $2.12 \pm 0.30$ \\
1342180815 & 07/27.782 & 2.039 & 1.850 & 4b & FSw & 1113 & 2802 & $0.994 \pm 0.023$ &             $+42\,\pm\,$ & 15 & $1.70 \pm 0.03$ \\